# Revisiting contrast mechanism of lateral piezoresponse force microscopy

Jaegyu Kim,[1,*] Seongwoo Cho,[1] Jiwon Yeom,[1] Seongmun Eom,[1] and Seungbum Hong[1,2,*]

[1]*Department of Materials Science and Engineering, Korea Advanced Institute of Science and Technology, Daejeon 34141, Republic of Korea*

[2]*KAIST Institute for the NanoCentury, Korea Advanced Institute of Science and Technology, Daejeon 34141, Republic of Korea*

Piezoresponse force microscopy (PFM) has been widely used for nanoscale analysis of piezoelectric properties and ferroelectric domains. Although PFM is useful because of its simple and nondestructive features, PFM measurements can be obscured by non-piezoelectric effects that could affect the PFM signals or lead to ferroelectric-like behaviors in non-ferroelectric materials. Many researches have addressed related technical issues, but they have primarily focused on vertical PFM. Here, we investigate significant discrepancies of lateral PFM signals between the trace and the retrace scans, which are proportional to the scan angle and the cantilever lateral tilting discrepancy. The discrepancies of PFM signals are analyzed based on intrinsic and extrinsic components, including out-of-plane piezoresponse, electrostatic force, and other factors. Our research will contribute to the accurate PFM measurements for visualization of ferroelectric in-plane polarization distributions.

*kimjaegyu@kaist.ac.kr, seungbum@kaist.ac.kr

Keywords: ferroelectric, polarization, piezoresponse force microscopy, cantilever lateral torsion, electrostatic force

# I. INTRODUCTION

Piezoresponse force microscopy (PFM), one of scanning probe microscopy (SPM) modes, is extensively used for nondestructive and nanoscale characterization of piezoelectric and ferroelectric materials. Piezoelectric properties and ferroelectric domains of materials with a nonzero piezoelectric tensor can be analyzed by the vibrational displacement (electromechanical response) of the cantilever-tip-sample system in PFM [1]. The displacement is induced by the converse piezoelectric effect (piezoresponse) driven by an external ac bias voltage. Based on the lock-in technique, vertical PFM (VPFM) [lateral PFM (LPFM)] visualizes out-of-plane (OP) [in-plane (IP)] domains using the longitudinal (torsional) vibration of the cantilever correlated with the effective piezoelectric coefficient $d_{33}^{eff}$ ($d_{15}^{eff}$). Furthermore, two-dimensional IP piezoresponse maps can be reconstructed by angle-resolved PFM (AR-PFM) based on LPFM results obtained at several specific angles between the long axis of the cantilever and the crystallographic axis of the sample [2,3]. Although PFM conveniently provides the polarization distributions using cantilever bending, it has typical technical issues related to cantilever dynamics [4], in addition to the resolution limit by the tip radius.

The PFM signals can be accompanied by non-piezoelectric effects such as the local electrostatic effect between the tip and the sample [5] and the non-local electrostatic effect between the cantilever and the sample [6], electrochemical strains (Vegard strains) [7], electrostrictions [8], flexoelectric effects [9], and Joule heating [10]. In extreme cases, the non-piezoelectric effects result in enhanced PFM amplitudes [10] or ferroelectric-like behaviors in non-ferroelectric materials [11]. The dependence of PFM on the cantilever-tip-sample system also leads to the involvement of other non-piezoelectric effects including cantilever buckling [12,13], sliding of the tip [14], adsorbates on the surface [15], topographic features [16], shape factor of the cantilever [17], and damping [18,19]. Not only the non-piezoelectric effects, but also the IP (OP) piezoresponse can distort the VPFM (LPFM) signals [12,13,20]. Many studies have addressed the cross-talk issues above [21-23] or suggested useful ways to utilize them [13]. However, there has been no study that fully addressed the coupling between OP and IP piezoresponse in LPFM signals.

In this study, we investigate the origin of discrepancies between the PFM signals of the trace scan (T) and those of the retrace scan (R) on ferroelectric polymer and oxide thin films. We visualize the discrepancies of the VPFM and LPFM signals and the lateral tilting (lateral friction) of the cantilever as a function of the scan angle. The discrepancies of the LPFM signals are proportional to the lateral tilting discrepancy proportional to the scan angle, whereas those of the VPFM signals does not significantly correlate with the lateral tilting discrepancy and the scan angle. In addition, their discrepancies are analyzed in terms of various factors, including the OP polarization state, non-local electrostatic force, lateral friction, PFM mode, and sample positioning. Based on the controlled experiments, we demonstrate the significant dependence of the LPFM discrepancy on the OP piezoresponse compared with the trivial non-local electrostatic force and the other factors. In some extreme cases, however, the non-local electrostatic force could be more dominant than the OP piezoresponse and the lateral tilting. Our study will be helpful in more accurate visualization of ferroelectric IP polarization distributions.

## II. EXPERIMENTAL PROCEDURE

SPM imaging on ferroelectric thin films was performed using an atomic force microscope (Cypher ES, Asylum Research) and two types of Pt/Ir-coated tips (EFM and CONTPt, Nanosensors) with spring constants of around 2.8 N/m and 0.2 N/m, respectively. The scan rate was 0.8 Hz and 1 Hz, and the loading force was between 50 nN and 150 nN. For ferroelectric thin films, we used poly(vinylidene fluoride-trifluoroethylene) [P(VDF-TrFE)] thin films (40 nm) on an Au bottom electrode on a Cr/SiO$_2$-deposited Si substrate [24], an BiFeO$_3$ (BFO) thin film (35 nm) epitaxially grown on a SrRuO$_3$ (75 nm)-deposited SrTiO$_3$ (001) substrate [25], and sol-gel processed polycrystalline Pb(Zr$_{0.2}$Ti$_{0.8}$)O$_3$ and Pb(Zr$_{0.52}$Ti$_{0.48}$)O$_3$ (PZT) thin films (50 nm and 100 nm) on a Pt (150 nm)/Ti (10 nm)/SiO$_2$-deposited Si substrate (Quintess Co. Ltd., Korea). We applied an ac modulation voltage of 1 V for the P(VDF-TrFE) and BFO thin films and 0.5 V for the PZT thin films to the tip while grounding the bottom electrode in PFM imaging, including vector PFM performed near contact resonant frequencies of around 380 kHz for VPFM and 700 kHz for LPFM, off-resonance vector PFM performed at contact frequencies of 17 kHz for VPFM and 37 kHz for LPFM, and dual ac resonance tracking LPFM (DART-LPFM) tracking with the contact resonance near 687 kHz. Lateral force microscopy (LFM) was performed simultaneously with PFM imaging.

## III. SIGNIFICANT DISCREPANCIES IN LATERAL PIEZORESPONSE FORCE MICROSCOPY SIGNALS BETWEEN TRACE AND RETRACE SCANS

### A. Scan angle dependence

To investigate the dependence of PFM signals on the scan direction, we first created domain structures in the P(VDF-TrFE) thin film by scanning an area of 4 μm × 4 μm using a grounded tip, while applying dc bias voltage steps of -10 V, -8 V, -6 V, -4 V, +4 V, +6 V, +8 V, and +10 V ($V_{BE}$) over a scan area of a width of 4 μm and a length of 500 nm to the bottom electrode. Figs. 1(a) and 1(b) show topography, VPFM, LPFM, and cantilever lateral torsion images on the electrically poled domain structure obtained with a scan angle ($\theta$) of 0° and 90° where the topography and the lateral torsion was simultaneously measured with PFM. Upper row images and lower row images were obtained with the trace scan and the retrace scan.

The P(VDF-TrFE) thin film had lamellar structures, which represents the feature of ferroelectric phase of P(VDF-TrFE), without any morphological damage after electric poling. Since P(VDF-TrFE) has negative $d_{33}^{eff}$ and $d_{15}^{eff}$ coefficients [26-28], which affect OP and IP piezoresponse in PFM, the VPFM phase signal was 0° on the up domain or 180° on the down domain, and the LPFM phase signal was 0° on the left-pointing IP domain or 180° on the right-pointing IP domain with respect to the cantilever long axis [24,29,30]. The OP polarization was switched to downward below -8 V, but upward above +6 V. It is intriguing that the upward switching was easier than downward switching when given the fact the pristine state was the down polarization. Both the VPFM and LPFM amplitude signals were large inside up- and down-poled domains, but they were relatively small in the region of +6 V of the up domain, which may be due to the unsaturated switching or the dead layer between the P(VDF-TrEF) thin film and the bottom electrode [31]. Large VPFM and LPFM amplitude signals indicate that the IP polarization may result from (110) polarization that includes both the OP and IP polarization components [32].

More interestingly, the LPFM phase signals had a remarkable opposite trend between the trace scan and the retrace scan only at the 90° scan angle, while the VPFM amplitude and phase signals had similar values between the trace scan and the retrace scan regardless of the scan angle. For instance, in the trace scan with the 90° scan angle compared to the case of the 0° scan angle, the pristine region and the down-poled domain had lower LPFM phase signals (darker violet phase), but the up-poled domain had higher LPFM phase signals (brighter yellow phase). However, the trend was reversed in the retrace scan with the 90° scan angle. Note that we define this difference between the trace scan and the retrace scan as "discrepancy", which is calculated by subtracting the retrace scan images from the trace scan images (T-R) [33].

The discrepancy maps [Figs. 1(c) and 1(d)] more clearly demonstrate the difference between the trace scan and the retrace scan. The topography discrepancy had negligible values, which indicates that the cantilever drift during PFM scanning was not the main factor for the large LPFM discrepancies. The VPFM and LPFM amplitude discrepancies were large mainly inside the poled domains, whereas the VPFM and LPFM phase discrepancies were large primarily at the domain boundaries, and the LPFM discrepancies were significantly larger than the VPFM discrepancies at the 90° scan angle. Usually, the VPFM signals tend to be more susceptible to cantilever buckling, which is maximum at the 0° scan angle, than cantilever lateral torsion, which is maximum at the 90° scan angle [13]. In contrast, the LPFM signals seem to be more susceptible to the cantilever lateral torsion.

It should be noted that the LPFM amplitude discrepancy depended on both the polarities of OP and IP polarizations. The LPFM amplitude discrepancy in the left-pointing IP domains was positive in the down-poled domain, but negative in the up-poled domain, and the trend was reversed in the right-pointing IP domains. The LPFM phase discrepancy primarily depended on the polarity of OP polarization. The LPFM phase discrepancy was mainly negative at the IP domain boundaries in the down-poled domain, but mainly positive at the IP domain boundaries in the up-poled domain.

The lateral torsion discrepancy represents the signal of transverse shear microscopy (TSM) at the 0° scan angle and the signal of LFM, which is also called friction force microscopy, at the 90° scan angle. In this paper, the lateral torsion discrepancy is used instead of TSM and LFM for easier comparison and uniform terminology. The lateral torsion discrepancy at both scan angles show only the morphological features of P(VDF-TrFE) lamellar structures. It was also considerably larger at the 90° scan angle than at the 0° scan angle, which suggests a strong correlation between the LPFM discrepancies and the lateral torsion discrepancy. However, the lateral torsion discrepancy did not have any feature of OP and IP domain structures at both scan angles, which means that the transverse shear or the lateral friction did not affect the large LPFM discrepancies.

**B. Possible mechanisms related to cantilever lateral torsion**
For better understanding of the principles of SPM scanning, we show schematic illustrations of the movements of a sample, a stage, and a cantilever as a function of the scan angle as shown in Fig. 2(a). In SPM scanning, the stage supporting the sample is moving in the xy-plane, but the tip-cantilever-chip system, which is fixed to the cantilever holder, is stationary in the xy-plane and moves in the z-axis. Although the cantilever does not move in the xy-plane, we can define its effective direction of movement in the opposite direction of the moving stage. The SPM tip, which is mounted on the end of the cantilever, scans one line back and forth along the fast scan direction with the trace (first) scan and the retrace (second) scan. After scanning the line, it moves a very short distance along the slow scan direction, which is the imaging direction, perpendicular to the fast scan direction and scans the next line with the trace and the retrace scans. The scan angle of 0° is when the fast scan direction of the cantilever is parallel to the long axis of the cantilever, and the scan angle of 90° is when the fast scan direction of the cantilever is perpendicular to the long axis of the cantilever.

Fig. 2(b) shows schematic illustrations of the moving sample and the tilting cantilever as a function of the scan angle. In contact modes, such as PFM and LFM, the cantilever tilts vertically (buckling) and/or laterally (torsion) in the opposite direction of the sample motion, since the tip contacting with the sample surface acts as a pivot due to the lateral friction, and its lateral torsion is the largest when the scan angle is 90°. In more detail, the higher the scan angle is for up to 90°, the greater the cantilever laterally tilts, where it tilts to the right (+y-direction) in the trace scan and tilts to the left (-y-direction) in the retrace scan. As the scan angle increases in the constant deflection mode [34], which is the mode for topography in PFM, the difference of the average lateral deflection between the trace scan and the retrace scan (lateral torsion discrepancy) increases, while the difference of the average vertical (longitudinal) deflection between the trace scan and the retrace scan is theoretically constant.

In LPFM scanning, the cantilever laterally oscillates clockwise or anticlockwise by the converse piezoelectric effect of the IP polarization excited by the sinusoidal ac modulation voltage. At the 90° scan angle, the cantilever tilts to the right with the positive average lateral deflection in the trace scan and tilts to the left with the negative average lateral deflection in the retrace scan. As a result, the cantilever oscillates at a laterally tilted position in the opposite directions in the trace scan and the retrace scan. Fig. 2(c) shows schematic illustrations of cantilever lateral oscillation under a downward electric field pointing from the tip to the bottom electrode in LPFM scanning with the 90° scan angle as a function of the driving force for the cantilever lateral oscillation. The LPFM phase reflected by the cantilever lateral oscillation is in-phase when the driving force is IP piezoresponse, whereas it would be out-of-phase when the driving forces are the OP components, i.e. the OP piezoresponse and the non-local electrostatic force, or other issues, such as the lateral friction, the contact resonance shift, and the lateral lag. In more detail, under the downward electric field, the cantilever laterally tilts clockwise to the right (+y direction) on right-pointing (+y direction) IP domains due to the shear strain ($d_{15}^{eff}$) both in the trace scan and the retrace scan (in-phase), and the LPFM amplitude and phase would be similar, which leads to negligible LPFM discrepancy. On the other hand, the cantilever laterally tilts clockwise to the right (+y direction) in the trace scan or anticlockwise to the left (-y direction) in the retrace scan (out-of-phase) on a down domain due to the vertical strain ($d_{33}^{eff}$), since the tip contacting with the sample surface acts as a pivot. Similarly, on the down domain, the cantilever laterally tilts clockwise to the right (+y direction) in the trace scan or anticlockwise to the left (-y direction) in the retrace scan (out-of-phase) due to the non-local electrostatic force [35]. The other factors, such as the lateral friction, the contact resonance shift, and the lateral lag also could induce the cantilever lateral tilting in the opposite directions in the trace scan and the retrace scan (out-of-phase).

**C. Detailed analysis of dependence of lateral piezoresponse force microscopy signals on cantilever lateral torsion**

In order to investigate the effect of the cantilever lateral torsion on the PFM discrepancy, PFM was conducted on the same region of Fig. 1 with different the scan angle from -90° to 90° with an angle step of 15° as shown in Fig. S1, and the corresponding discrepancy maps were calculated as shown in Fig. S2. We did not sequentially conduct PFM with the scan angle from -90º to 90º to exclude the effects of experimental factors, such as cantilever tuning, humidity, and temperature.

Fig. 3(a) shows the lateral torsion discrepancy as a function of the scan angle extracted from Fig. S2. The higher the scan angle was, the greater the lateral torsion discrepancy between the trace scan and the retrace scan as discussed in the previous section. This indicates the frictional isotropy in the P(VDF-TrFE) thin film regardless of the domain structures [36]. The lateral torsion discrepancy was positive or negative when the scan angle was positive or negative, since in the trace scan the cantilever moves to the right (+y direction) with the positive scan angles or moves to the left (-y direction) with the negative scan angles.

Figs. S1 and S2 demonstrate the more detailed trend of VPFM and LPFM discrepancies as a function of the lateral torsion discrepancy. Contrary to trivial VPFM discrepancies, LPFM discrepancies were proportional to the lateral torsion discrepancy. To be specific, as the scan angle increased, the LPFM phase primarily increased in the up-poled domain or decreased in the down-poled domain and the pristine region, when the cantilever was laterally tilted to the right (+y direction), which was the case of the trace scan with the positive scan angles or the retrace scan with the negative scan angles. On the other hand, the trend was reversed when the cantilever was laterally tilted to the left (-y direction).

Figs. 3(b)–3(g) show the histograms of the discrepancy maps (Fig. S2) for quantitative analysis. As the scan angle increases, the LPFM amplitude discrepancy peak becomes smaller and broader, while the VPFM amplitude discrepancy peak was slightly shifts to the negative values. Likewise, as the scan angle increases, the LPFM phase discrepancy peak becomes smaller, and its two shoulders from 60° to 360° and -60° to -360° increases proportionally with the scan angle, but the VPFM phase discrepancy peak is almost the same. These results imply that the LPFM discrepancy is proportional to the cantilever lateral torsion rather than the cantilever buckling [13].

Histogram and Gaussian fitting results more clearly show the dependence of the LPFM phase on the OP polarization and the scan direction as shown in Figs. S3 and S4. The two peaks of the LPFM phase near 0° and 180° changes asymmetrically as functions of the scan angle and the OP polarization. As the scan angle increases on the up domain, in the trace scan, the peak near 0° almost does not shift and becomes slightly sharper, but the peak near 180° gradually shifts to the lower value and becomes broader. On the other hand, in the retrace scan, the peak near 0° gradually shifts to the lower value and becomes broader, but the peak near 180° almost does not shift and becomes sharper. On the down domain, the trends of the LPFM phase change are reversed. This indicates the strong dependence of the LPFM phase discrepancy on the OP piezoresponse in the P(VDF-TrFE) thin film, which can be explained within the framework of the cantilever lateral motion induced by the OP piezoresponse [Fig. 3(c)].

**D. Comparison of contributions from out-of-plane piezoresponse and non-local electrostatic force**

To explore the contributions of the OP components, such as the OP piezoresponse and the non-local electrostatic force, to the LPFM discrepancy [Fig. 2(c)], we conducted PFM with the 90° scan angle on box-patterned up- and down-poled domains in ferroelectric oxide and polymer thin films. Fig. 4 shows VPFM and LPFM discrepancy maps on the box-patterned domains in PZT thin films and a P(VDF-TrFE) thin film. Both the oxide and polymer thin films had the box-patterned LPFM discrepancies, which were much larger for the polymer thin film as shown in Gaussian fitting of the LPFM phase on up- and down-poled domains (see Fig. S6). This indicates that the LPFM discrepancy might be ubiquitous for various materials, of which magnitude depends on the properties of the material, e.g. piezoelectric constant, dielectric constant, elastic modulus, electromechanical coupling factor, and mechanical damping. However, we found that the lateral torsion and its discrepancies do not have any box-patterned features in all samples as shown in Fig S3, which means that the lateral friction would not be the main factor for the LPFM discrepancy.

As discussed in the previous section, the LPFM phase discrepancy at the IP domain boundaries strongly correlate with the polarity of OP polarization and the signs of $d_{33}^{eff}$ and $d_{15}^{eff}$ coefficients. For instance, it was primarily positive in the down-poled domain in the PZT thin films with positive $d_{33}^{eff}$ and $d_{15}^{eff}$ coefficients but mainly negative in the down-poled domain in the P(VDF-TrFE) film with negative $d_{33}^{eff}$ and $d_{15}^{eff}$ coefficients [26]. The opposite trend of the LPFM phase discrepancy implies that the LPFM phase discrepancy primarily arises from the OP piezoresponse that is dependent on the polarity of OP polarization rather than the non-local electrostatic force that are independent of the polarity of symmetric OP polarization. Therefore, applying dc bias voltage ($V_{dc}$) to the tip could not alleviate the LPFM discrepancy, since the OP piezoresponse is weakly dependent of $V_{dc}$, whereas the non-local electrostatic force is strongly dependent on $V_{dc}$.

It is worth noting that the LPFM discrepancies appeared even in the OP domains with very small VPFM amplitude signals (outer boxes) in the 50 nm PZT thin films. The small VPFM amplitude signals may be because the OP piezoresponse was offset by the nontrivial non-local electrostatic force. In fact, the OP piezoresponse of the outer boxes would affect the LPFM signals similarly to that of the inner boxes with large VPFM amplitude signals. The reason why the LPFM discrepancies were not proportional to the VPFM amplitude signals in Figs. S1 and S2 may be the same as in this case.

To further analyze the dependence of LPFM signals on the OP piezoresponse and the non-local electrostatic force, we employed a flexible cantilever with a low spring constant of ~0.13 N/m to maximize the electrostatic effects, since the flexible cantilever is strongly subjected to the non-local electrostatic effects [5,37]. Fig. 5(a) shows PFM results with 0° and 180° VPFM phase signals on box-patterned up- and down-poled domains in the P(VDF-TrFE) thin film visualized by a cantilever with a spring constant of ~2.71 N/m. In contrast, the VPFM phase signal obtained using the flexible cantilever in Fig. 5(b) was almost all 180° without the domain boundary feature, which means that the flexible cantilever was significantly subjected to the electrostatic effects .

Although the PFM phase signal indicated down polarization in the scanned area, the LPFM phase discrepancy was primarily negative in the down-poled region and slightly positive in the up-poled region, which is consistent with the previous results in Figs. 1d and 4d. This supports that the LPFM phase discrepancy is more sensitive to the OP polarization than the electrostatic force. In addition, we adjusted the electrostatic force by applying dc bias voltage ($V_{dc}$) to the tip during PFM scanning on the box-patterned domains. Continuous scanning at the center line of box-patterned domains while applying 0 V, +2 V, +4 V, -2 V, and -4 V to the tip in sequence significantly changed the VPFM amplitude signal probably because of the electrostatic effects linearly depend on ($V_{CPD}$-$V_{dc}$), where $V_{CPD}$ is the contact potential difference between the tip-cantilever system and the sample [6] [Fig. 5(c)]. During line scanning, the OP polarization was partially switched by $V_{dc}$ of +4 V and -4 V. Despite the significant change in the VPFM amplitude signal by a factor of up to 15, the LPFM discrepancies were not reversed or significantly changed. This implies that the LPFM discrepancies primarily depend on the OP polarization rather than the non-local electrostatic force.

We also adjusted the electrostatic effects on the cantilever with the spring constant of ~2.71 N/m used in Fig. 5(a) by applying 0 V, -4 V, and +4 V to the tip during PFM scanning [Figs. 5(d)-5(f)]. The VPFM signals were strongly affected by the tip bias. The VPFM amplitude signal was enhanced in the up-poled domain by the tip bias of -4 V and in the down-poled domain by the tip bias of +4 V, respectively. The VPFM phase signal represents the dominant up domain or down domain and partial upward switching by the tip bias of -4 V and partial downward switching by the tip bias of +4 V. The LPFM phase discrepancy was sensitive to the partial switching of OP polarization, which was manifested by its partial change with respect to the OP polarization, rather than the electrostatic effects. The up or down OP polarization can be represented by primarily positive (blue) or negative (red) boundary area in the LPFM phase discrepancy in the P(VDF-TrFE) film. This new marker for OP polarization is expected to provide detailed distinctions of complex distributions of OP polarization.

If the OP piezoresponse is a dominant factor, it would enhance or hinder the cantilever lateral oscillation as a function of the scan direction based on a simple hypothesis. Fig. 6 shows hypothetical illustrations of the cantilever lateral oscillation driven by the OP piezoresponse and the IP piezoresponse. For example, the cantilever laterally tilts clockwise to the right (0° LPFM phase) by the IP piezoresponse in both the trace and the retrace scans under a downward electric field on a down-poled, right-pointing IP domain with the 90° scan angle in a ferroelectric thin film with positive $d_{33}^{eff}$ and $d_{15}^{eff}$ coefficients [Fig. 6(a)]. At this time, the clockwise cantilever lateral torsion would be enhanced in the trace scan because the cantilever additionally tilts clockwise as the OP piezoresponse lifts the tip, a pivot point of the cantilever-tip-sample system, which may result in larger LPFM amplitude and lower LPFM phase. On the other hand, the clockwise cantilever lateral torsion would be hindered in the retrace scan because the cantilever tilts anticlockwise due to the OP piezoresponse lifting the tip, which may lead to smaller LPFM amplitude and higher LPFM phase [see Fig. 5(a)]. As a result, the LPFM amplitude and phase discrepancies on the down-poled, right-pointing IP domain would be positive and negative, respectively.

Table 1 summarizes the LPFM signals from the hypothesis of the dominant OP piezoresponse and the experiment based on the histograms. Compared to the hypothesis, the experimental results of LPFM phase on the IP domains are partially correct, and those of LPFM amplitude are all in the opposite trend. The origins of opposite results are unclear at this stage and still remain as an open question, but at least the OP piezoresponse appears to be more dominant than the non-local electrostatic force because the LPFM phase at the IP domain boundaries depends on the OP piezoresponse.

**E. Possibility of engagement of non-local electrostatic force**

Does the non-local electrostatic force play no role at all in the LPFM discrepancy? We chose an BFO thin film with down polarization to exclude the effect of the polarity of OP polarization. Fig. 7 show LPFM signals and lateral torsion on the BFO thin film before and after sample rotation by 90° clockwise ($\phi$). The LPFM discrepancies were proportional to the lateral torsion discrepancy regardless of the sample position, which suggests that the LPFM discrepancy would not mainly result from the PFM scanning conditions, such as inclination of the sample relative to the cantilever lateral torsion. The LPFM amplitude discrepancy mainly came from inside the IP domains and was positive in the left-pointing IP domain and negative in the right-pointing IP domain with respect to the cantilever long axis as presented in Fig. 1. Almost all positive LPFM phase discrepancy primarily occurred at the IP domain boundaries because the BFO thin film had the down polarization.

Note the 90° scan angle case. The retrace scan, compared to the trace scan, had smaller LPFM amplitude signals in the left-pointing (-y direction) IP domains and larger ones in the right-pointing (+y direction) IP domains, which leads to obscure IP domain boundaries. In addition, the right-pointing IP domains were much larger in the retrace scan, and they could cover the entire area in more extreme case. The more detailed dependence of LPFM signals on the scan angle and the lateral torsion are represented in Fig. S7. This phenomenon is similar to the case where the VPFM amplitude signals are ambiguous at OP domain boundaries and the VPFM phase signals have no phase inversion between up and down domains due to the significant non-local electrostatic force between the cantilever and the sample [6,37]. In other words, the non-local electrostatic force may induce the LPFM discrepancy to some extent.

**F. Significant non-local electrostatic force**

In the previous section, we discussed the extreme case where the entire area is almost covered by one IP domain possibly due to the extremely strong non-local electrostatic force. To verify the extreme case, PFM was conducted at the edges of a PZT thin film with down polarization as in the case of the BFO thin film as shown in Fig. 8. The non-local electrostatic force between the cantilever and the sample was symmetric [Fig. 8(a)] or asymmetric [Figs. 8(b) and 8(c)] with respect to the long axis of the cantilever. In the symmetric case, the non-local electrostatic force may have a similarly trivial effect on the LPFM signals in the trace and the retrace scans since the long axis of the cantilever is perpendicular to the sample edge. On the other hand, in the asymmetric case, the electrostatic force may affect the LPFM signals in the opposite directions in the trace and the retrace scans since the long axis of the cantilever is parallel to the sample edge. Fig. S8 shows the corresponding optical microscope images of the cantilever and the sample.

In the symmetric case, the significant LPFM discrepancy only occurred when the scan angle was 90° as discussed so far. In the asymmetric cases, however, the LPFM phase discrepancy was significantly governed by the sample position with respect to the cantilever regardless of the cantilever lateral torsion, the scan angle, and the scan direction. When the sample was placed on the left side of the cantilever, the LPFM phase signals primarily represented the left-pointing (-y direction) IP domain, i.e. IP domain with IP polarization pointing towards the sample [Fig. 8(b)], and the LPFM amplitude signals were larger in the trace scan where the cantilever was laterally tilted far from the sample. On the other hand, when the sample was placed on the right side of the cantilever [Fig. 8(c)], the LPFM phase signals primarily exhibited the right-pointing (+y direction) IP domain, i.e. IP domain with IP polarization pointing towards the sample, in any edge of the sample, and the LPFM amplitude signals were larger in the retrace scan where the cantilever laterally was tilted far from the sample. These results indicate that the non-local electrostatic force could have a more significant effect on the LPFM signals than the OP piezoresponse in some extreme cases. The reverse trend of the LPFM amplitude signals is clearly presented in the LPFM amplitude discrepancy maps.

It should be noted that in all the asymmetric cases, the LPFM amplitude signals were larger and the LPFM phase signals were more unidirectional when the cantilever was laterally tilted far from the sample than when it was laterally tilted close to the sample. This may be due to the competition between the non-local electrostatic forces in the opposite directions. For example, when the cantilever is at the bottom edge of the sample [Fig. 8(b)], the LPFM signals would be affected by the non-local electrostatic forces in the -y and +y directions and the IP piezoresponse. Since the sample is placed on the left side of the cantilever (more negative y-axis value), the magnitude of the non-local electrostatic force in the -y direction would be much greater than that of the non-local electrostatic force in the +y direction, and the non-local electrostatic force in the +y direction may be very weak when the cantilever is laterally tilted far from the sample. Therefore, the non-local electrostatic force in the -y direction would be dominant when the cantilever is laterally tilted far from the sample (trace scan) or less dominant when the cantilever is laterally tilted close to the sample (retrace scan). In other words, the LPFM signals would be affected by the electrostatic force in the -y direction dominantly in the trace scan or less dominantly in the retrace scan. As a result, LPFM results in the trace scan would have the larger amplitude signals and the more unidirectional phase signals due to the dominant non-local electrostatic force in the -y direction, but LPFM results in the retrace scan would have the relatively smaller amplitude signals and the relatively less unidirectional phase signals due to the less dominant non-local electrostatic force in the -y direction and the involvement of the relatively weaker IP piezoresponse.

**G. Piezoresponse force microscopy modes**

It has been shown that the OP components, such as the OP piezoresponse and the non-local electrostatic force, could significantly affect the LPFM signals, but the lateral friction does not have a correlation with the LPFM discrepancy. To verify the other factors, such as the contact resonance of oscillation of the cantilever-tip-sample system and the lateral lag, PFM was conducted in the off-resonance PFM mode and the DART-LPFM mode on up- and down-poled domains in the P(VDF-TrFE) thin film as shown in Fig. S9. The domains were created by the same procedure in Fig. 1. The DART-LPFM mode uses two amplitudes at two drive frequencies offset by a certain frequency to a frequency lower (drive frequency1) or higher (drive frequency2) than the contact resonant frequency and tracks the contact resonance shift by shifting the drive frequencies until the difference of the two amplitudes is zero [38]. The drive frequency in the off-resonance PFM mode was 17 kHz for VPFM and 37 kHz for LPFM and that in the DART-LPFM mode was around 687 kHz. Here, the lateral torsion discrepancy had no domain features (Fig. S9c), which means that the lateral friction is not the main factor for the large LPFM discrepancies.

Fig. 9 shows the corresponding discrepancy maps of VPFM and LPFM. The off-resonance PFM modes and the DART-LPFM mode did not have significant discrepancy at the 0° scan angle, except the amplitude discrepancies in the off-resonance VPFM mode and the DART-LPFM mode, which may be due to the large cantilever vertical torsion (buckling) [13]. In addition, the off-resonance VPFM mode did not have large discrepancies when the scan angle was 90°. However, both the off-resonance LPFM mode that does not use the contact resonance and the DART-LPFM mode that tracks the contact resonance shift exhibited significant LPFM discrepancies when the scan angle was 90°. These results support that the contact resonance is not the main factor for the LPFM discrepancy. In addition, it is noteworthy that in the off-resonance PFM, only at the 90° scan angle, the LPFM amplitude was larger in the down-poled domain where the VPFM amplitude signals were larger than in the up-poled domain (Fig. S9a). This implies the involvement of the OP piezoresponse in the off-resonance LPFM signals.

For easier comparison, Table 2 summarizes the key factors which contribute to the LPFM signals and the LPFM discrepancies in the trace and retrace scans at the 90° scan angle. First, the IP piezoresponse causes the cantilever lateral torsion in the same direction by the shear strain ($d_{15}$) leading to the negligible LPFM discrepancy regardless of the scan direction. Second, the OP piezoresponse would result in the cantilever lateral torsion in the opposite directions by the vertical strain ($d_{33}$), which has a major effect on the LPFM signals and the LPFM discrepancies. Third, the non-local electrostatic force between the cantilever and the sample would cause the cantilever lateral torsion in the opposite directions by the electric forces in the opposite directions, which would be smaller than the effect of the OP piezoresponse but much more critical than the effect in some extreme cases such as sample edge scanning. The other factors such as the lateral friction, PFM mode, and lateral lag would cause the mechanically induced signal delays, which would have a trivial effect on the LPFM signals. Since the LPFM discrepancy is severe at the larger scan angle, we recommend the 0° scan angle for LPFM scanning in order to exclude it and carefully analyze the LPFM results obtained with both the trace and the retrace scans.

## IV. CONCLUSION

In conclusion, we have investigated the significant discrepancies of LPFM signals between the trace and the retrace scans that increase with increasing lateral torsion of the cantilever, regardless of the PFM mode. The discrepancies in the LPFM signals were much more severe than those in the VPFM signals, and they were ubiquitous in various types of ferroelectric materials such as oxides and polymers. Through controlled experiments, we found that the LPFM discrepancy primarily resulted from the OP piezoresponse but could be dominated by the non-local electrostatic force between the cantilever and the sample in some extreme cases. Our study will provide insights into the nanoscale in-plane surface movement and enable more accurate visualization of ferroelectric domains.

## ACKNOWLEDGMENTS

This research was conducted as part of the KAIST-funded Global Singularity Research Program for 2021 and 2022 under award number 1711100689, and the Ministry of Education of the Republic of Korea and the National Research Foundation of Korea (NRF-2019S1A5C2A03081332).

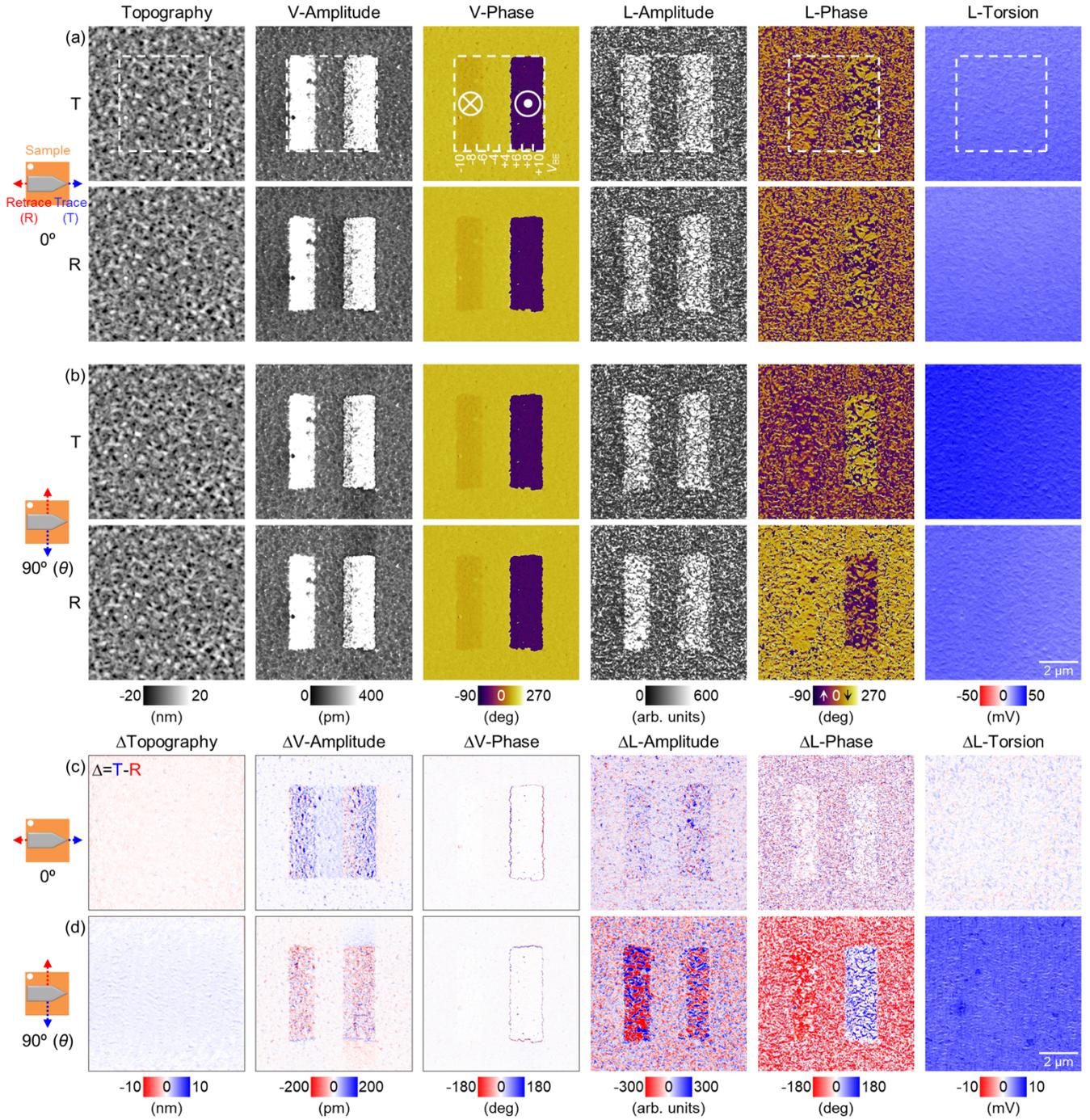

FIG 1. Topography, vertical and lateral PFM (VPFM and LPFM), and lateral torsion images on up- and down-poled domains in a P(VDF-TrFE) thin film obtained with a scan angle ($\theta$) of 0° (a) and 90° (b) where the fast scan direction of the cantilever in trace (T) and retrace (R) is indicated by blue and red dotted arrows. The lateral torsion (L-Torsion) was measured simultaneously with PFM. The up and down domains were created by scanning a 4×4 μm² area (denoted by dash frames) from left to right using a grounded tip while applying -10 V, -8 V, -6 V, -4 V, 4 V, 6 V, 8 V, and 10 V ($V_{BE}$) at 500 nm intervals to the bottom electrode. Discrepancy (Δ) maps of topography, VPFM, LPFM, and lateral torsion calculated by subtracting the retrace scan images from the trace scan images (T-R) obtained with a scan angle ($\theta$) of 0° (c) and 90° (d).

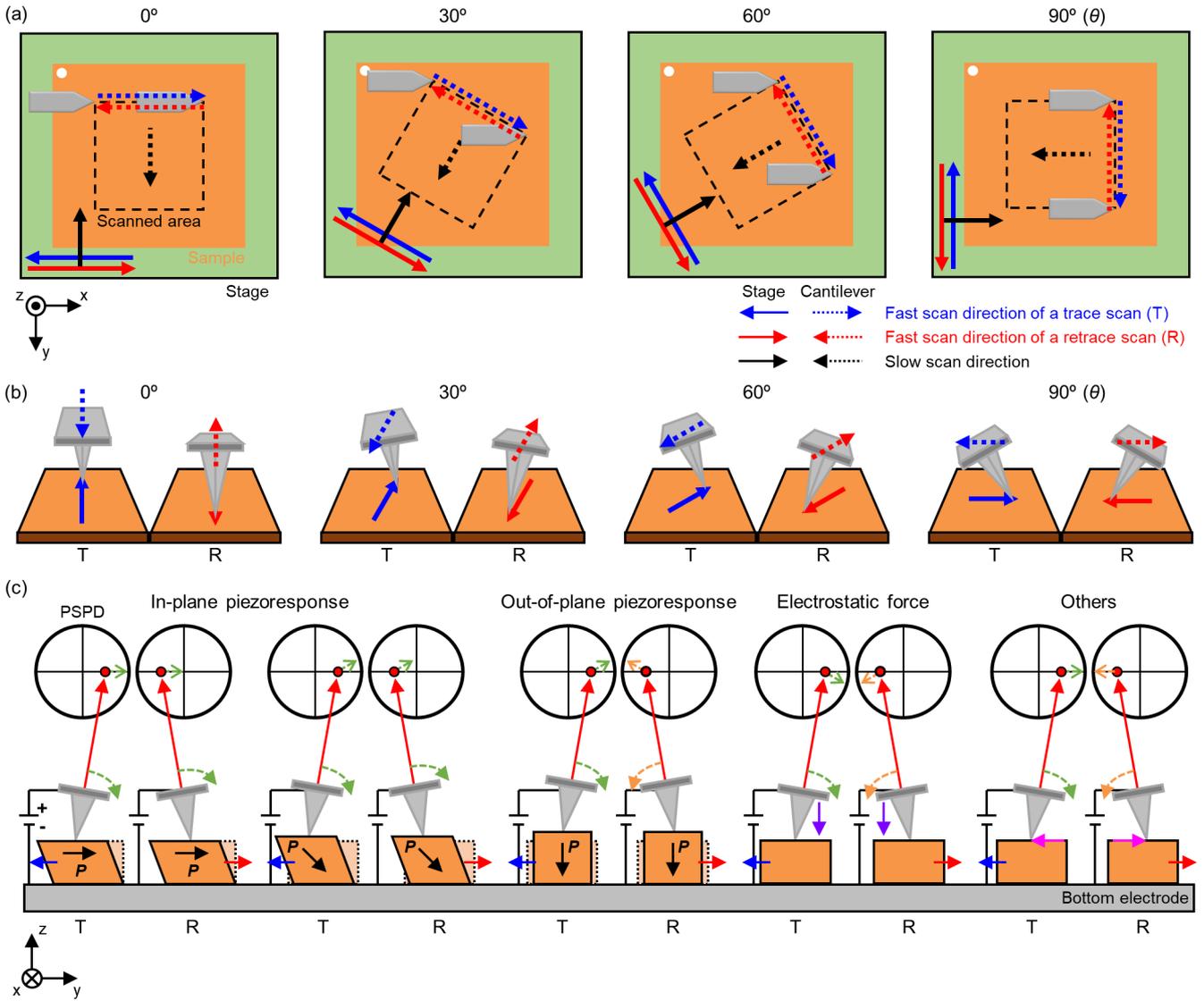

FIG 2. (a) Top-view and (b) front-view schematic illustrations of cantilever, sample and stage motion in PFM scanning where the stage on which the sample is fixed is moving, while the cantilever is fixed to the cantilever holder as a function of scan angle ($\theta$). The scan directions of the stage and the cantilever are indicated by solid and dotted arrows in blue, red, and black for the fast trace scan, the fast retrace scan, and the slow scan, respectively. The scan directions of the cantilever are opposite to those of the stage. The lateral torsion of the cantilever caused by the stage motion and its discrepancy between the trace scan and the retrace scan increase as the scan angle increases. (c) Rear-view schematic illustrations of cantilever lateral torsion, sample and stage motion in LPFM scanning with the scan angle of 90° as a function of the main factor of cantilever lateral torsion: in-plane piezoresponse, out-of-plane piezoresponse, non-local electrostatic force, and other issues, such as lateral friction, contact resonance shift, and lateral lag. P represents the ferroelectric polarization.

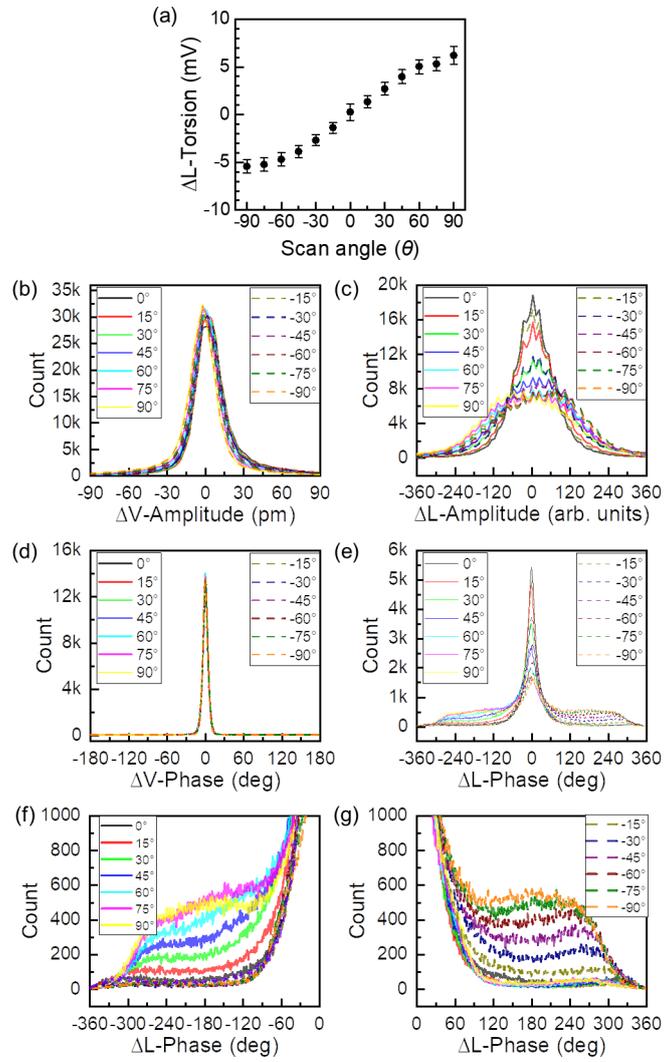

FIG 3. Discrepancy (Δ) graphs of lateral torsion (a), VPFM (b) and (d), and LPFM (c), (e), (f), and (g) calculated from Figs. S1 and S2 as a function of scan angle (θ). Insets in the graphs indicate the scan angles.

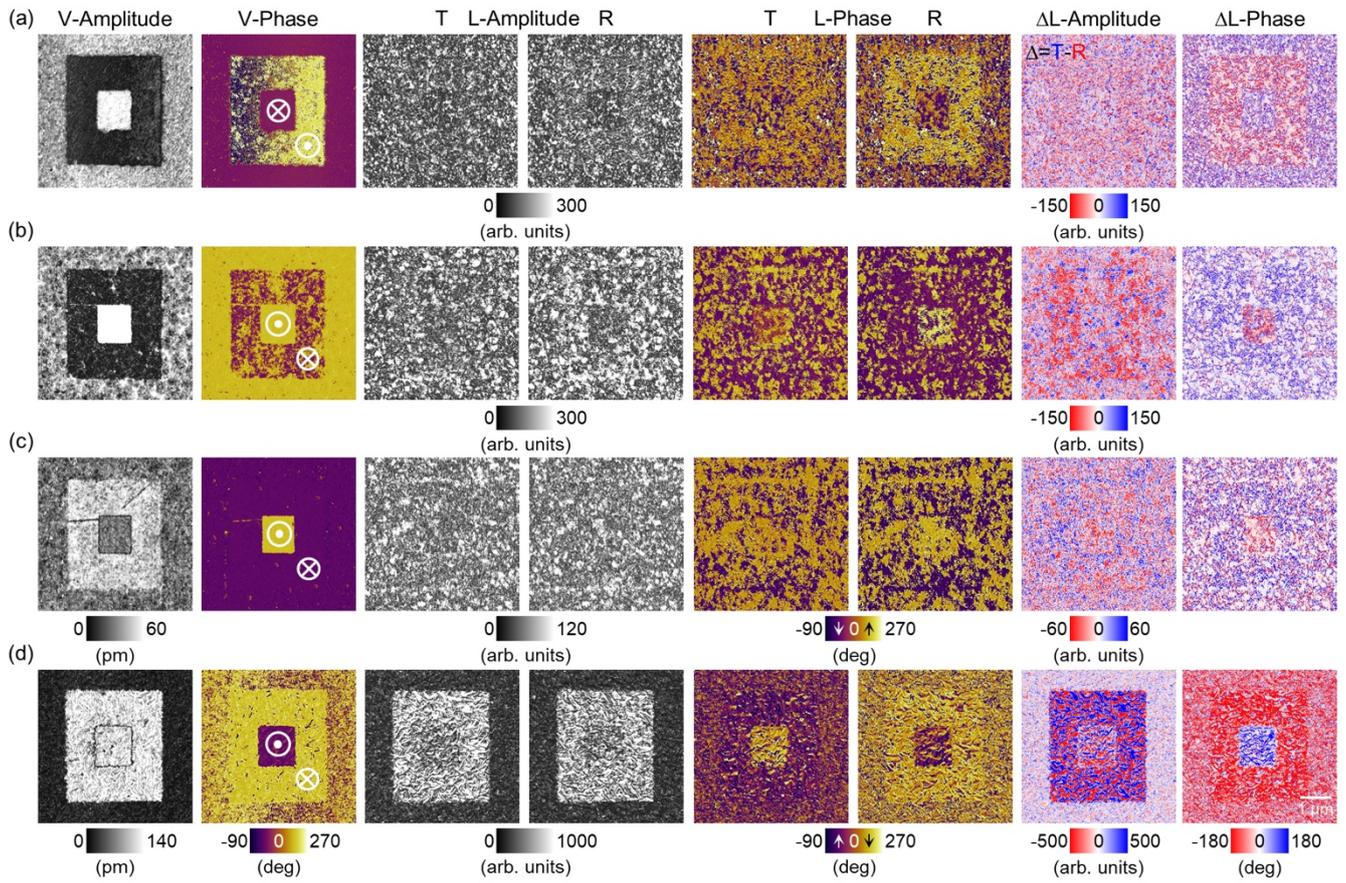

FIG 4. VPFM and LPFM images and discrepancy (Δ) maps of LPFM on box-patterned domains in ferroelectric oxide and polymer thin films obtained with a scan angle of 90°: (a) 50 nm PZT (20/80), (b) 50 nm PZT (52:48), (c) 100 nm PZT (52:48), and (d) 20 nm P(VDF-TrFE).

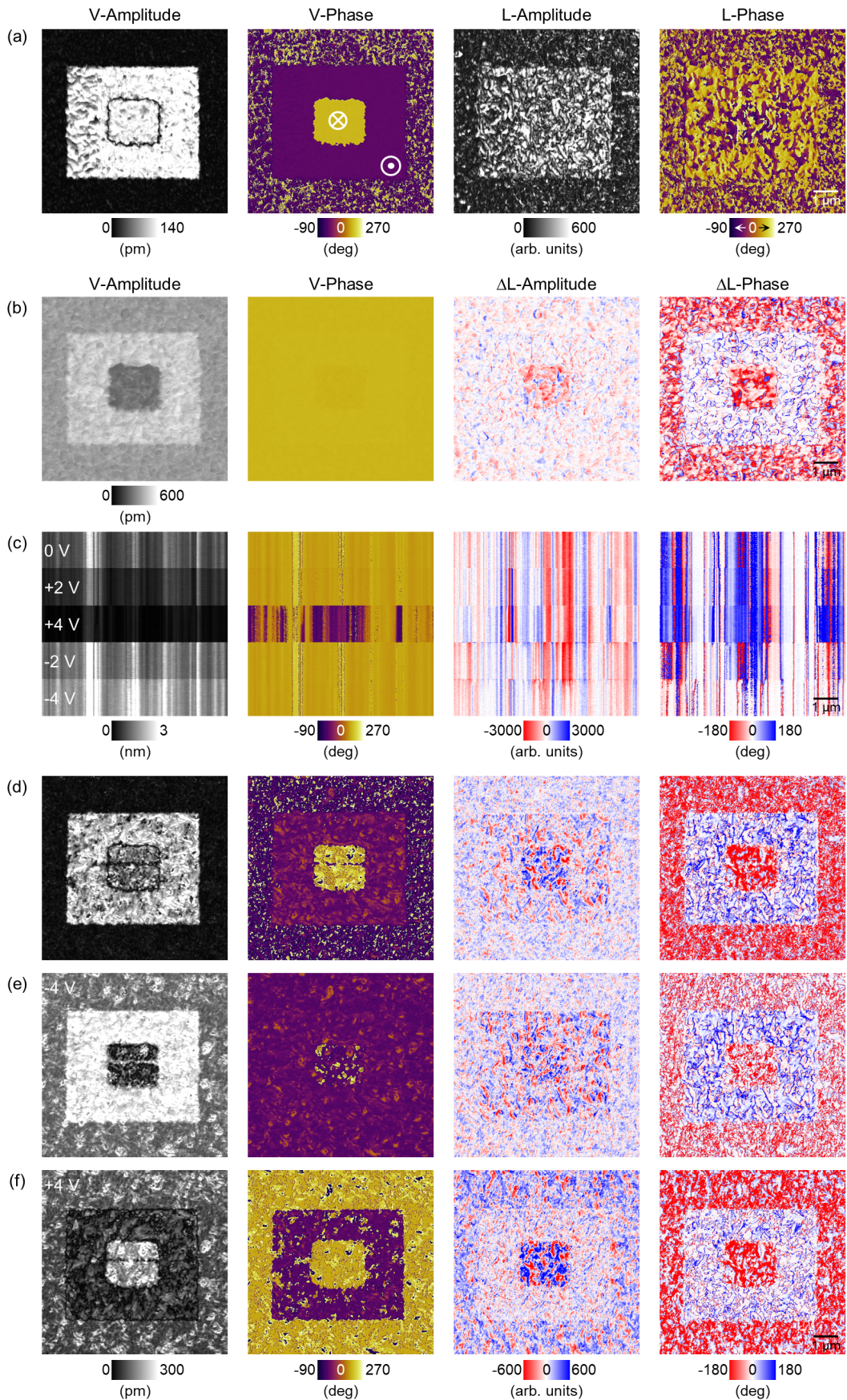

FIG 5. (a) VPFM and LPFM images on box-patterned domains in a P(VDF-TrFE) film obtained using a cantilever with a spring constant of ~2.71 N/m and with a scan angle of 90°. (b-c) VPFM images and LPFM discrepancy (Δ) maps obtained using a cantilever with a spring constant of ~0.13 N/m: (c) lines scan on the center of the box-patterned domains while applying dc bias voltage of 0 V, +2 V, +4 V, -2 V, and -4 V to the tip. (d-f) VPFM images and LPFM discrepancy (Δ) maps obtained using the cantilever used in (a) with a scan angle of 90° while applying 0 V (d), -4 V (e), and +4 V (f) to the tip.

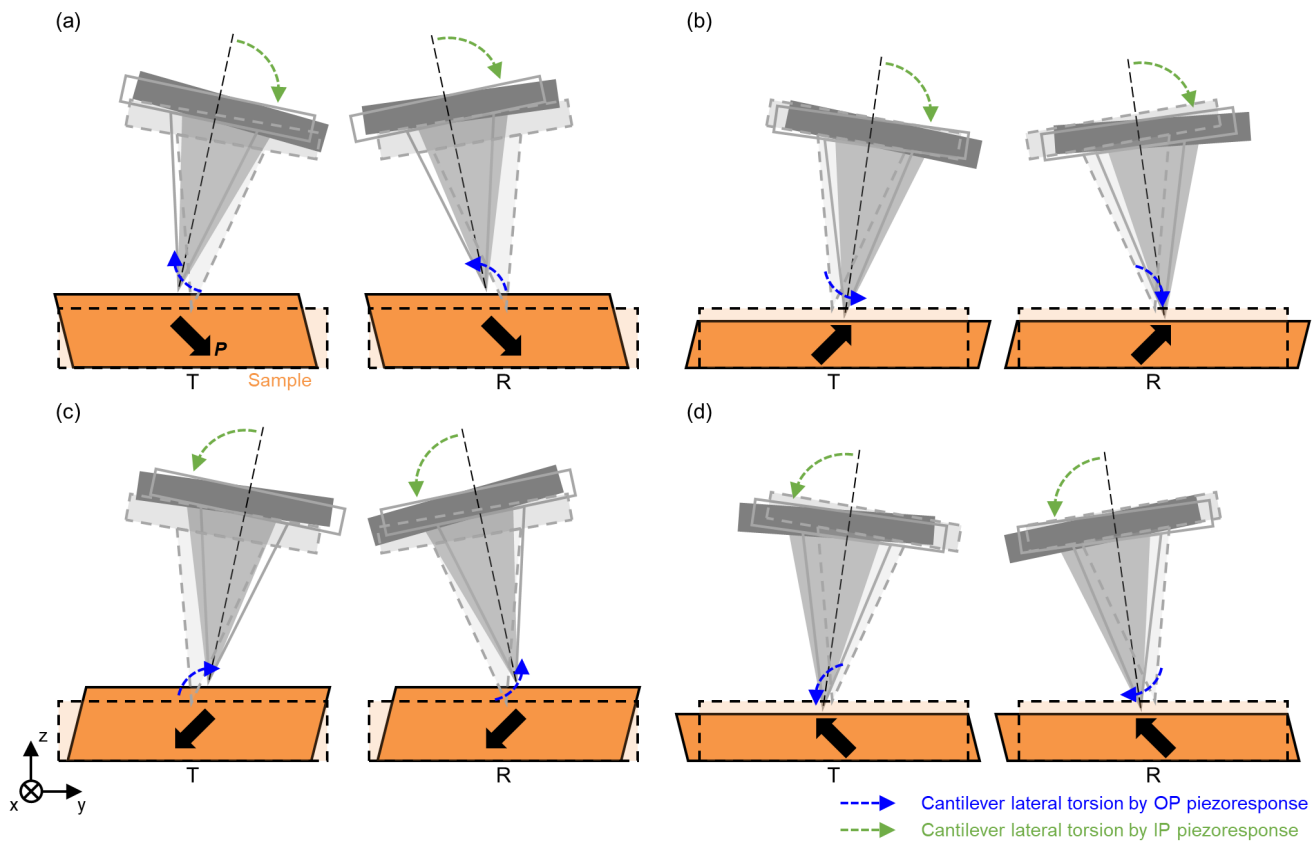

FIG 6. Schematic illustrations of cantilever and sample motion in PFM scanning as a function of polarization direction (a), (b), (c), and (d) in trace (T) and retrace (R) scans in a sample with positive $d_{33}^{eff}$ and $d_{15}^{eff}$ coefficients under a downward electric field or in a sample with negative $d_{33}^{eff}$ and $d_{15}^{eff}$ coefficients under an upward electric field.

TABLE 1. LPFM signals from the hypothesis of the dominant OP piezoresponse and the experiment

| Polarization direction | LPFM amplitude on IP domains $d_{33}^{eff}, d_{15}^{eff}$ | | | LPFM phase on IP domains $d_{33}^{eff}, d_{15}^{eff}$ | | | LPFM phase at IP domain boundaries $d_{33}^{eff}, d_{15}^{eff}$ | |
|---|---|---|---|---|---|---|---|---|
| | Hypothesis | Positive | Negative | Hypothesis | Positive | Negative | Positive | Negative |
| ↘: down / right | T > R | T < R | T < R | T < R | T > R | T < R | T > R | T < R |
| ↙: down / left | T < R | T > R | T > R | T > R | T < R | T > R | T > R | T < R |
| ↗: up / right | T < R | T > R | T > R | T > R | T > R | T > R | T < R | T > R |
| ↖: up / left | T > R | T < R | T < R | T < R | T < R | T < R | T < R | T > R |

T and R indicates trace and retrace scans.

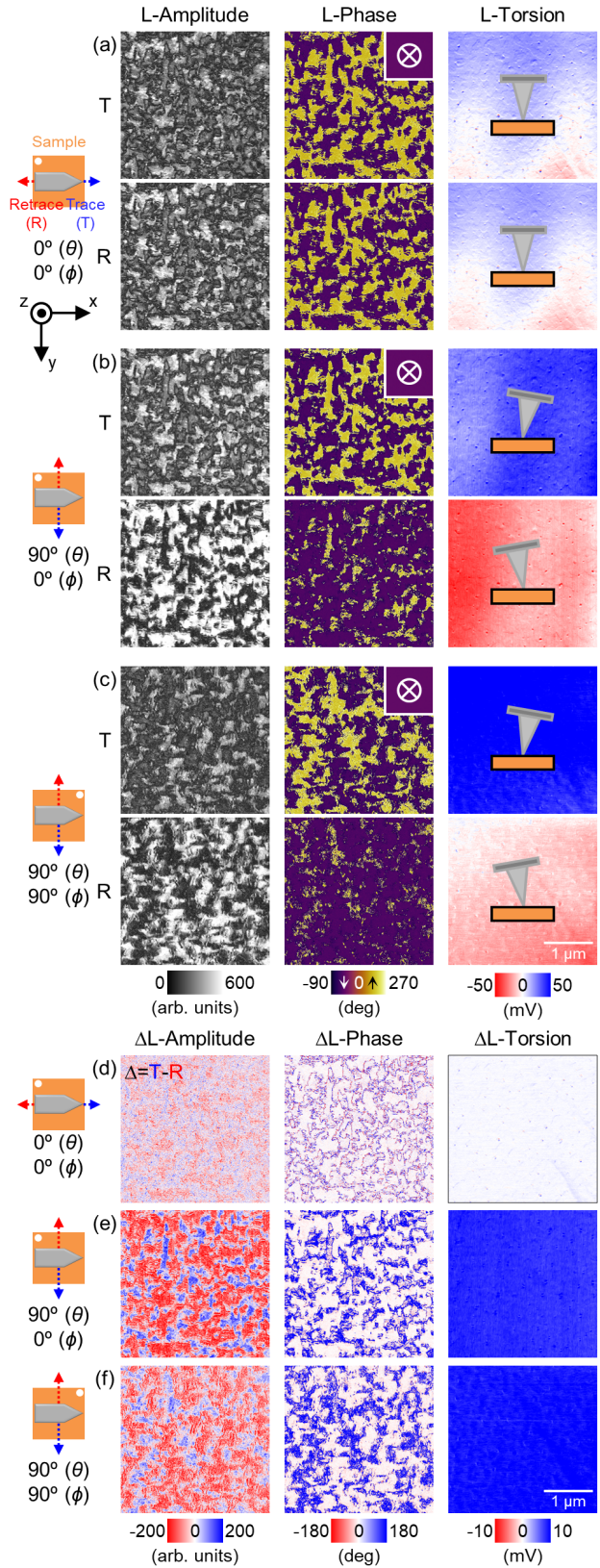

FIG 7. LPFM and lateral torsion images on a BFO thin film obtained with a scan angle ($\theta$) of 0° (a) and 90° (b) and (c) before (a) and (b) and after sample rotation ($\phi$) by 90° clockwise (c) where the fast scan direction of the cantilever in trace (T) and retrace (R) is indicated by blue and red dotted arrows. Inset: the corresponding VPFM phase images. The lateral torsion was measured simultaneously with PFM. Insets in the lateral torsion images represent the schematic illustrations of the cantilever lateral torsion and the sample in PFM scanning. (d), (e), and (f) Discrepancy ($\Delta$) maps of LPFM and the lateral torsion.

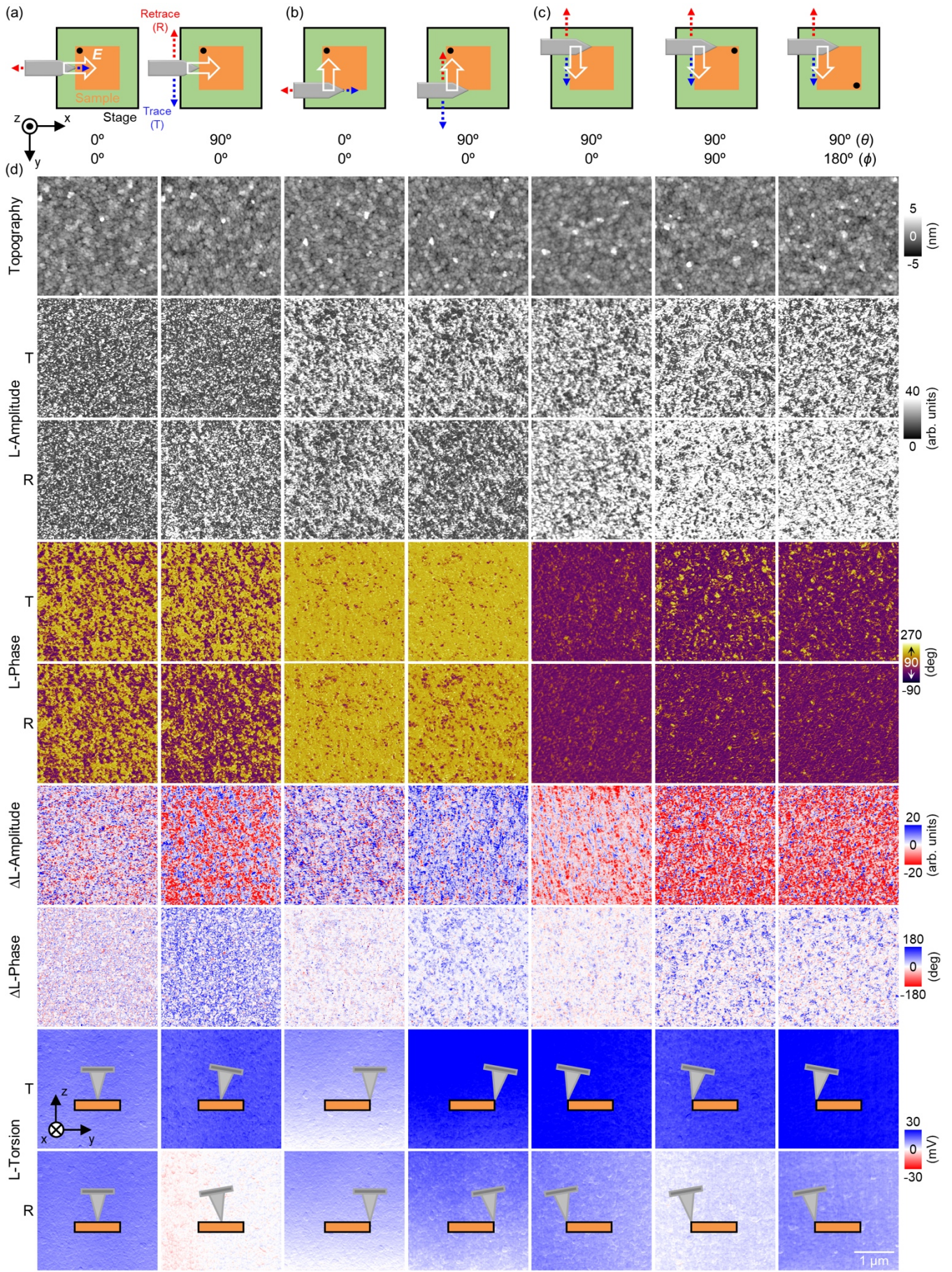

FIG 8. Schematic illustrations of cantilever motion in PFM scanning where the long axis of the cantilever is perpendicular (a) or parallel (b) and (c) to the sample edge, in which the fast scan direction of the cantilever in trace (T) and retrace (R) scans is indicated by blue and red dotted arrows as a function of scan angle ($\theta$). (d) Topography, LPFM, discrepancy ($\Delta$) maps of LPFM,

and lateral torsion images on a PZT thin film as a function of scan angle ($\theta$) and sample rotation angle ($\phi$). The direction of the non-local electrostatic force ($E$) between the cantilever and the sample is indicated by hollow arrows. The lateral torsion was measured simultaneously with PFM. Insets in the lateral torsion images represent the schematic illustrations of the cantilever lateral torsion and the sample in PFM scanning.

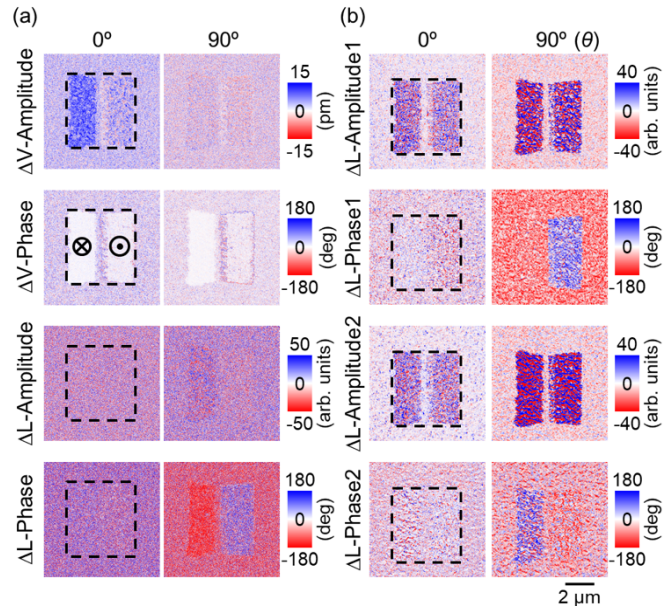

FIG 9. Discrepancy (Δ) maps of (a) off-resonance PFM mode and (b) DART-LPFM mode calculated by subtracting the retrace scan images from the trace scan images (T-R) obtained with a scan angle (θ) of 0° and 90° on up- and down-poled domains (denoted by dash frames) in a P(VDF-TrFE) thin film created by the same procedure in Fig. 1. Signals1 and signals2 in the DART-LPFM mode represent the signals obtained at the drive frequency1 and the drive frequency2, respectively. The discrepancy maps were calculated by subtracting the retrace scan images from the trace scan images (T-R).

TABLE 2. Key factors contributing to LPFM signals with 90° scan angle

| Driving force | Behavior in trace and retrace scans | Contribution to LPFM discrepancy |
| --- | --- | --- |
| IP piezoresponse | Same cantilever lateral torsion by the same shear strain of $d_{15}$ | X |
| OP piezoresponse | Opposite cantilever lateral torsion by the same vertical strain of $d_{33}$ | Major |
| Non-local electrostatic force | Opposite cantilever lateral torsion by the same electric force | Minor (Major at sample edge) |
| Others (Friction, PFM mode, Materials, etc.) | Mechanical signal delay | Minor |

**Supplemental Material**

# Revisiting contrast mechanism of lateral piezoresponse force microscopy

Jaegyu Kim,[1,*] Seongwoo Cho,[1] Jiwon Yeom,[1] Seongmun Eom,[1] and Seungbum Hong[1,2,*]

[1]*Department of Materials Science and Engineering, Korea Advanced Institute of Science and Technology, Daejeon 34141, Republic of Korea*

[2]*KAIST Institute for the NanoCentury, Korea Advanced Institute of Science and Technology, Daejeon 34141, Republic of Korea*

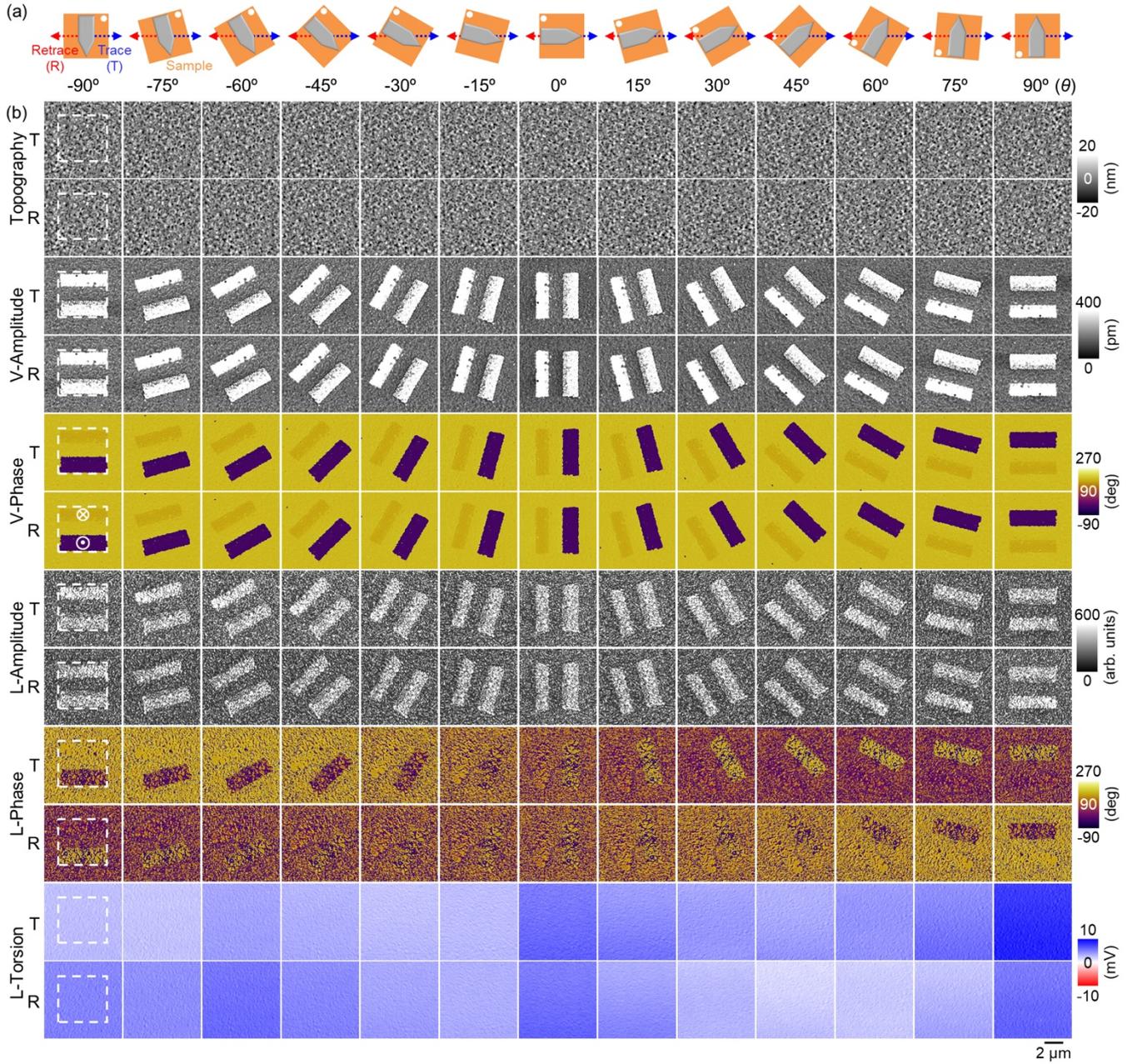

FIG S1. (a) Schematic illustrations of cantilever motion in PFM scanning where the fast scan direction of the cantilever in trace (T) and retrace (R) scans is indicated by blue and red dotted arrows as a function of scan angle ($\theta$). (b) Topography, vertical and lateral PFM (VPFM and LPFM), and lateral torsion (L-Torsion) images on up and down poled domains in a P(VDF-TrFE) thin film obtained with a scan angle ($\theta$) of 0° (a) and 90° (b) where the fast scan direction of the cantilever in trace (T) and retrace (R) is indicated by blue and red dotted arrows. The lateral torsion was measured simultaneously with PFM. The up and down domains were created by scanning a 4×4 μm$^2$ area (denoted by dash frames) from top to bottom using a grounded tip while applying -10 V, -8 V, -6 V, -4 V, 4 V, 6 V, 8 V, and 10 V at 500 nm intervals to the bottom electrode.

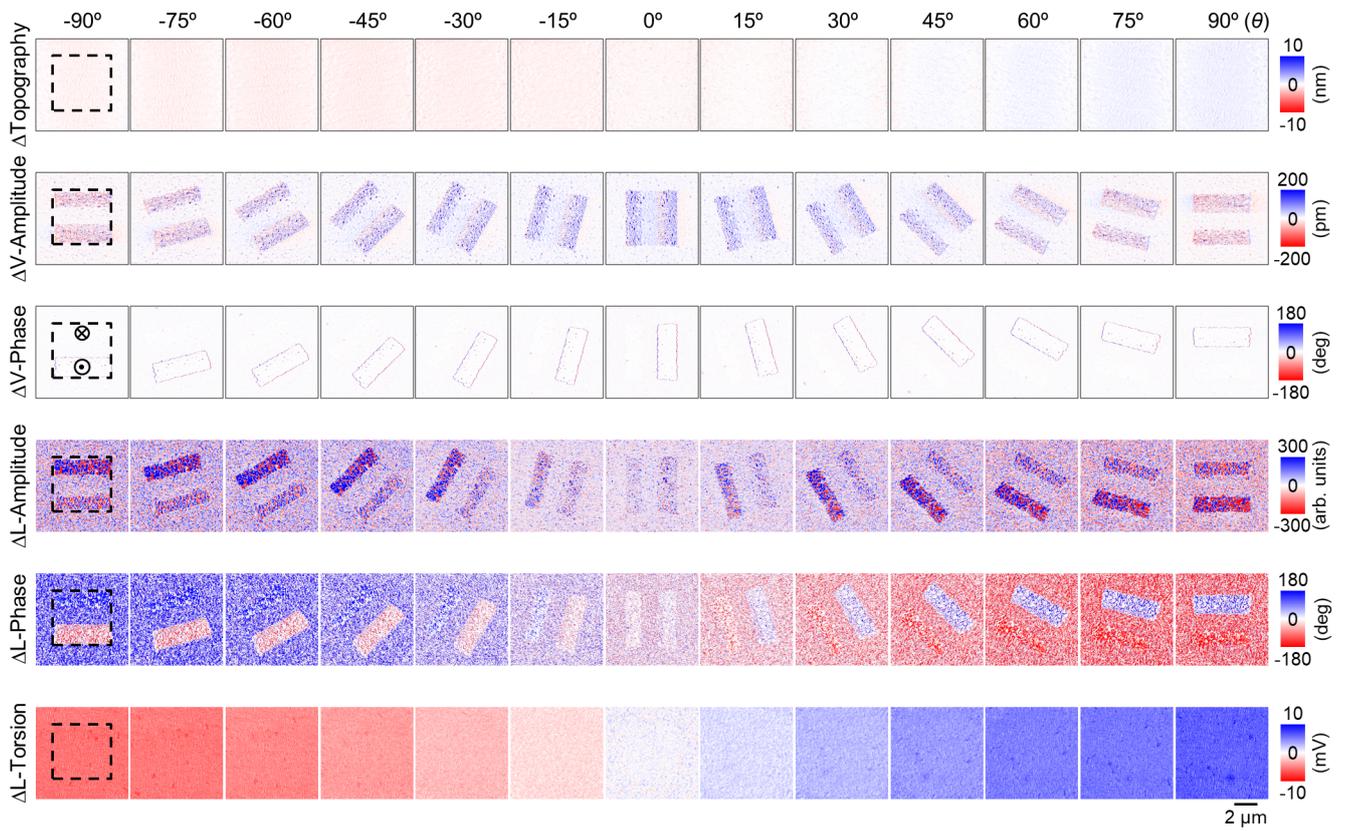

FIG S2. Discrepancy (Δ) maps of topography, VPFM, LPFM, and lateral torsion as a function of scan angle ($\theta$) of Fig. S1. The discrepancy maps were calculated by subtracting the retrace scan images from the trace scan images (T-R).

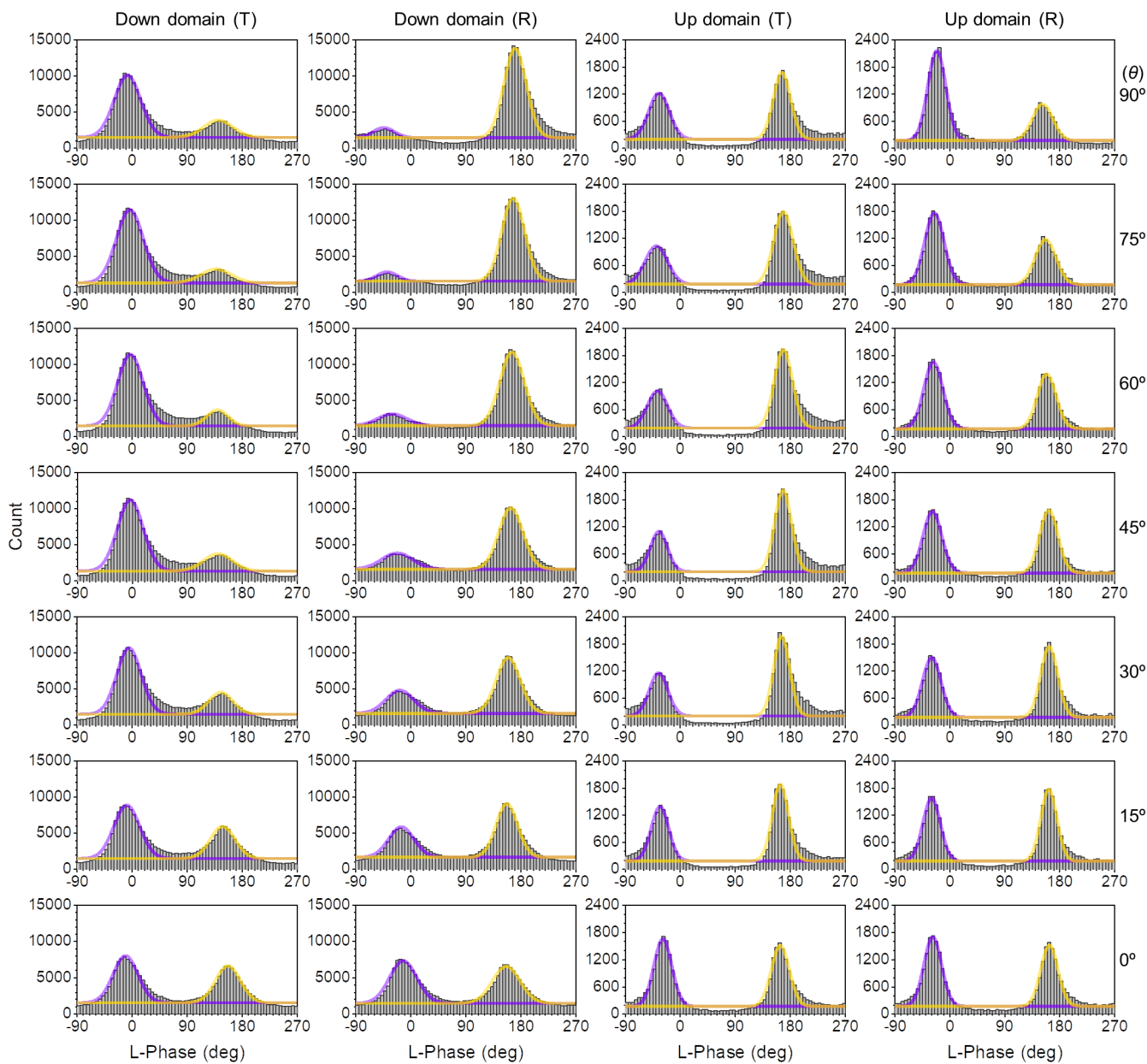

FIG S3. Histograms and calculated Gaussian fitting of LPFM phase in up and down domains in trace (T) and retrace (R) scans of Fig. S1 as a function of scan angle ($\theta$). The bin size is 5 for all graphs.

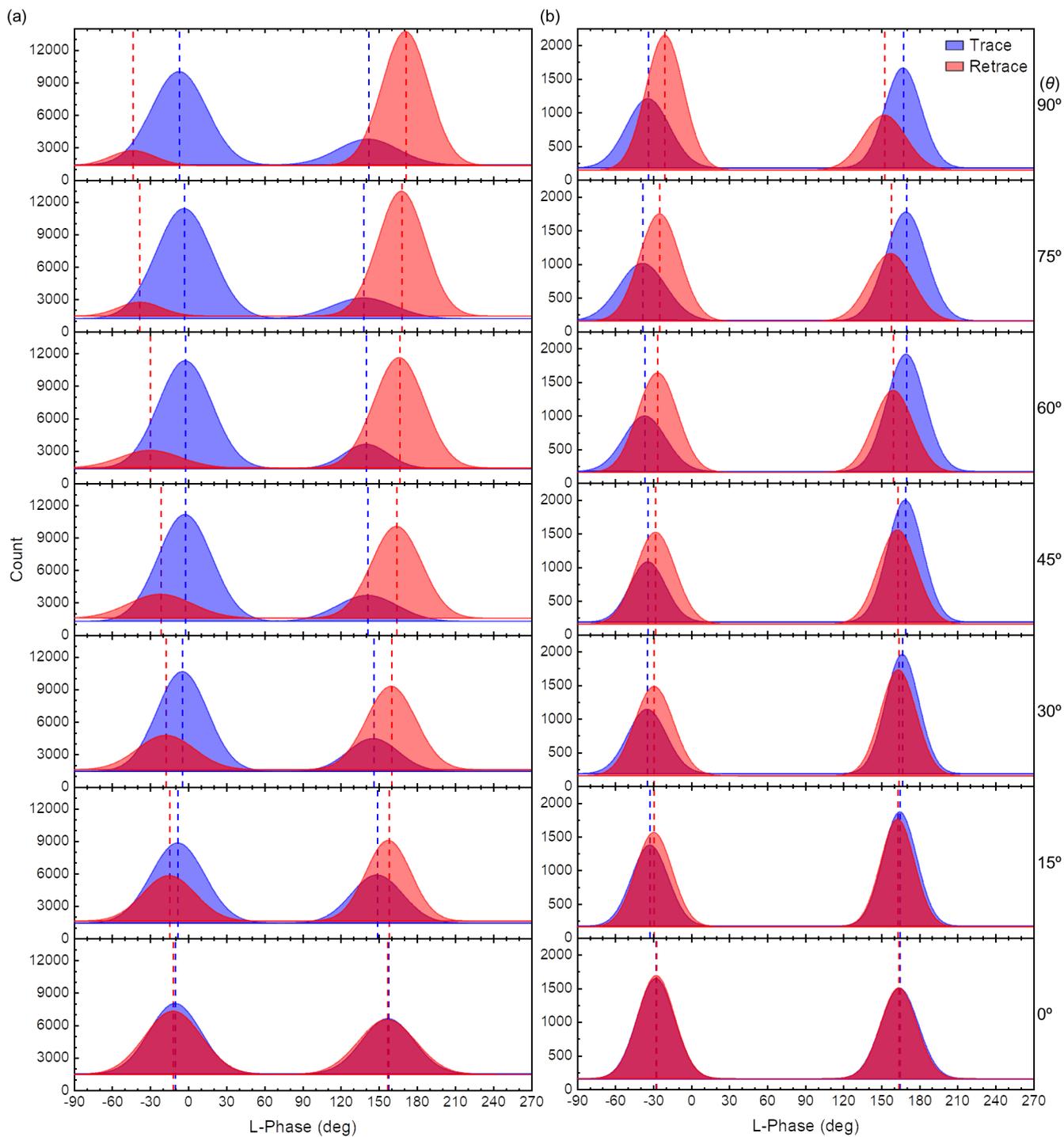

FIG S4. Gaussian fitting calculated from histograms of LPFM phase in up (a) and down (b) domains in trace and retrace scans of Fig. S3 as a function of scan angle ($\theta$).

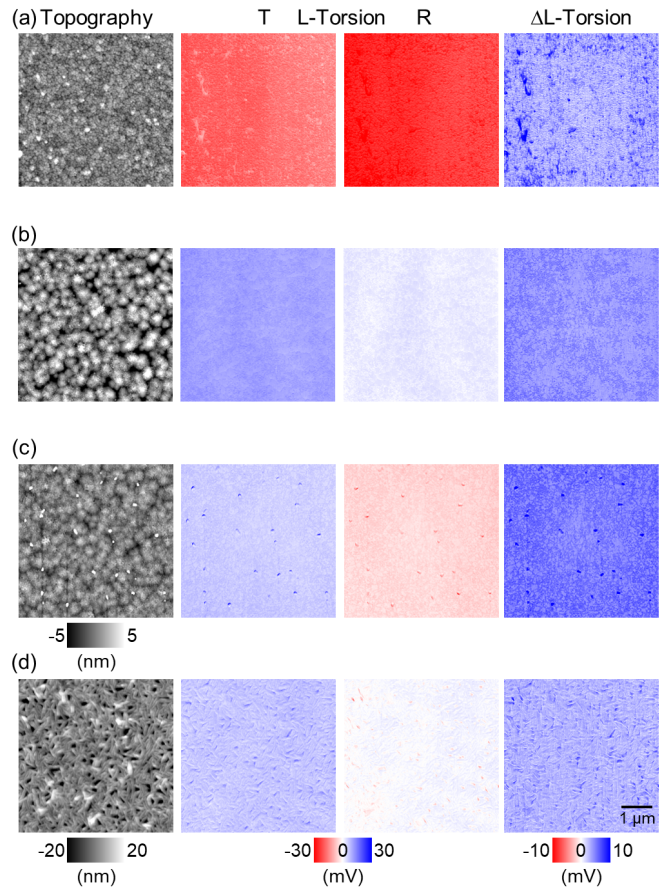

FIG S5. Topography and lateral torsion images and discrepancy (Δ) maps of lateral torsion on box-patterned domains in ferroelectric oxide and polymer thin films obtained with a scan angle of 90° of Fig. 4: (a) 50 nm PZT (20/80), (b) 50 nm PZT (52:48), (c) 100 nm PZT (52:48), and (d) 20 nm P(VDF-TrFE). The lateral torsion was measured simultaneously with PFM.

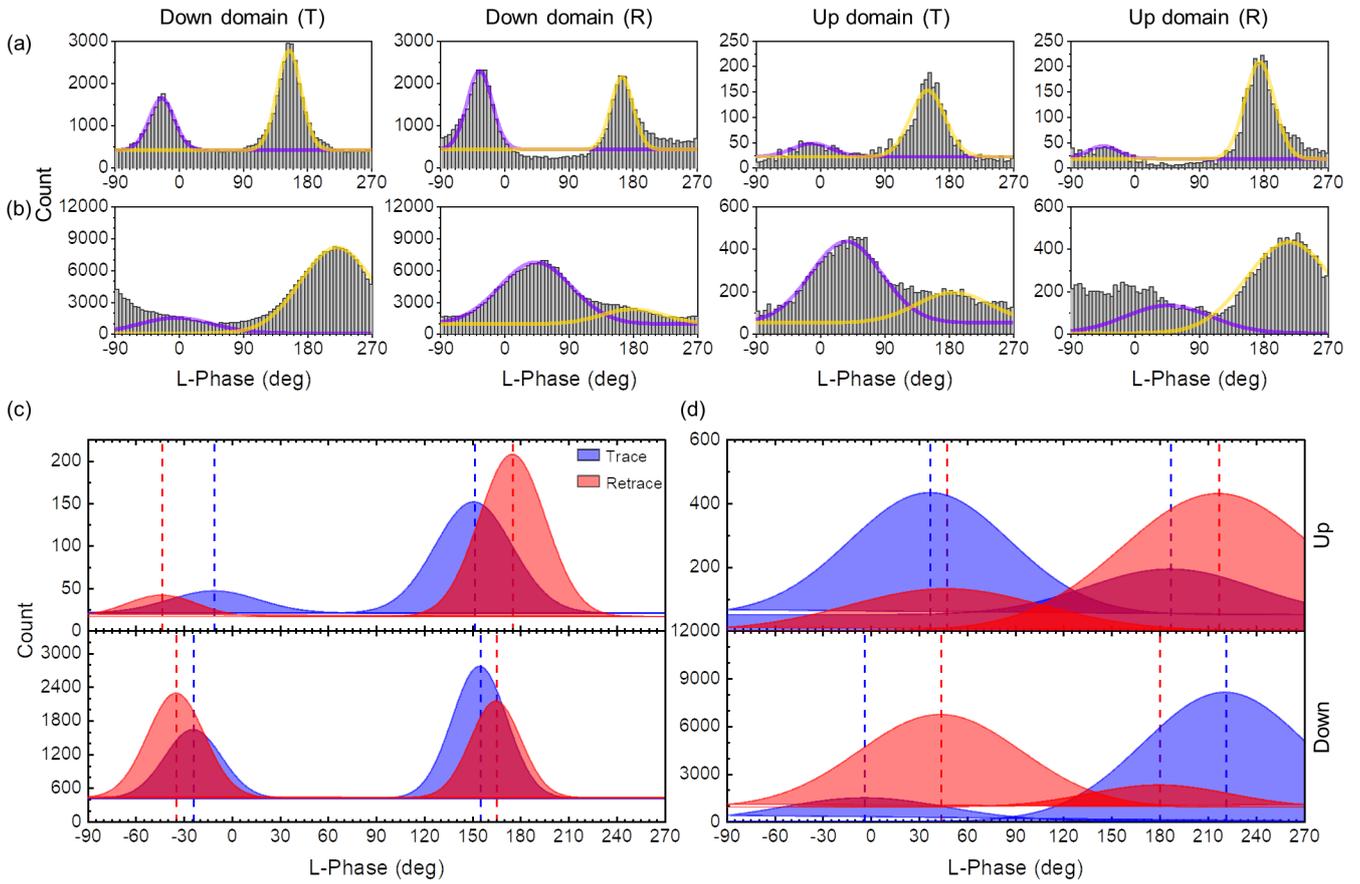

FIG S6. Histograms and calculated Gaussian fitting of LPFM phase in up and down domains in trace (T) and retrace (R) scans in 100 nm PZT (52:48) (a) and 20 nm P(VDF-TrFE) (b) thin films of Fig. 4. The bin size is 5 for all graphs. Gaussian fitting calculated from histograms of LPFM phase in up and down domains of 100 nm PZT (52:48) (c) and 20 nm P(VDF-TrFE) (d) thin films.

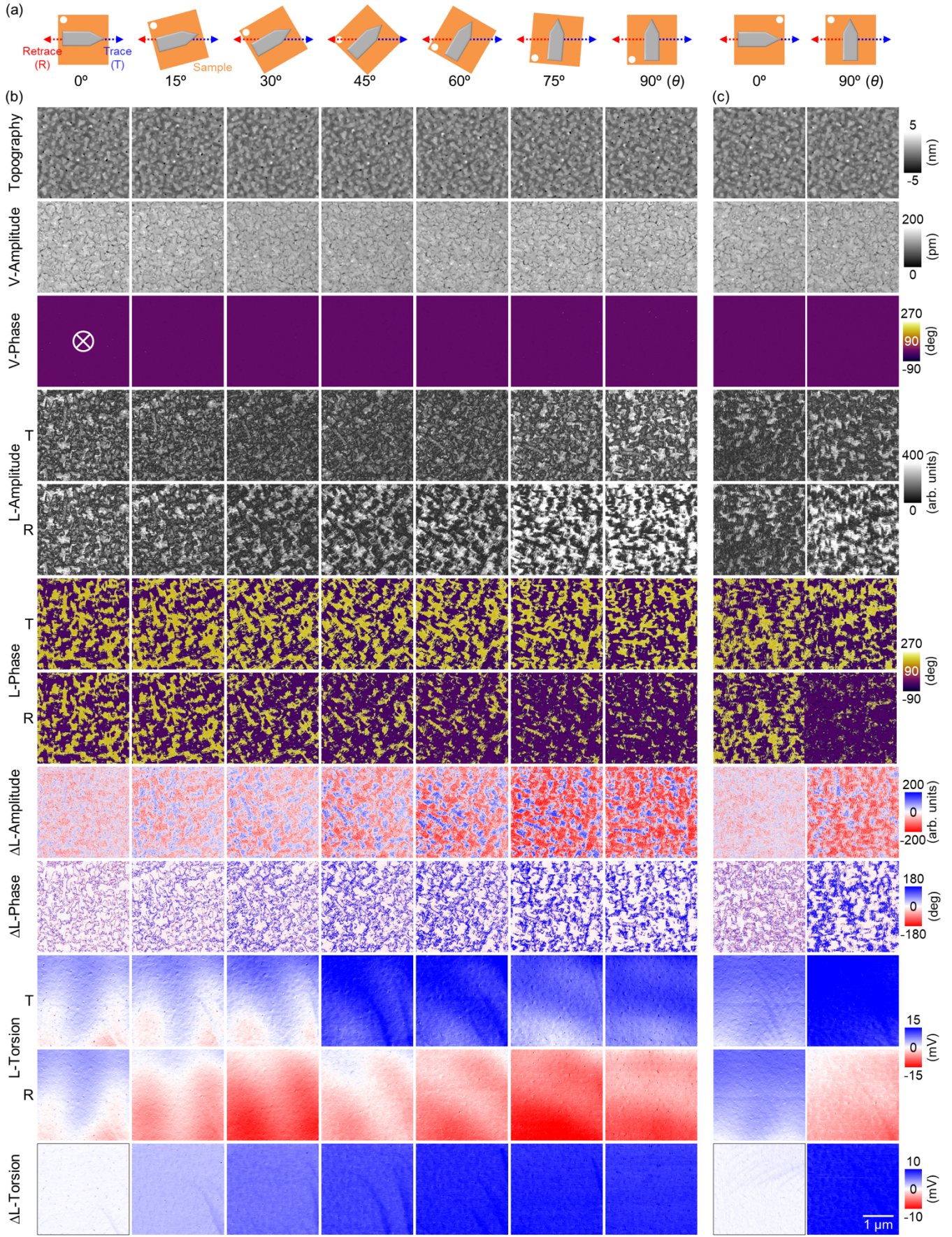

FIG S7. (a) Schematic illustrations of cantilever motion in PFM scanning where the fast scan direction of the cantilever in trace (T) and retrace (R) scans is indicated by blue and red dotted arrows as a function of scan angle ($\theta$). (b) Topography, VPFM, LPFM, and lateral torsion images and discrepancy ($\Delta$) maps of LPFM and lateral torsion on a BFO thin film (b) before and (c) after sample rotation ($\phi$) by 90° clockwise as a function of scan angle ($\theta$). The discrepancy maps were calculated by subtracting

the retrace scan images from the trace scan images (T-R). The lateral torsion was measured simultaneously with PFM.

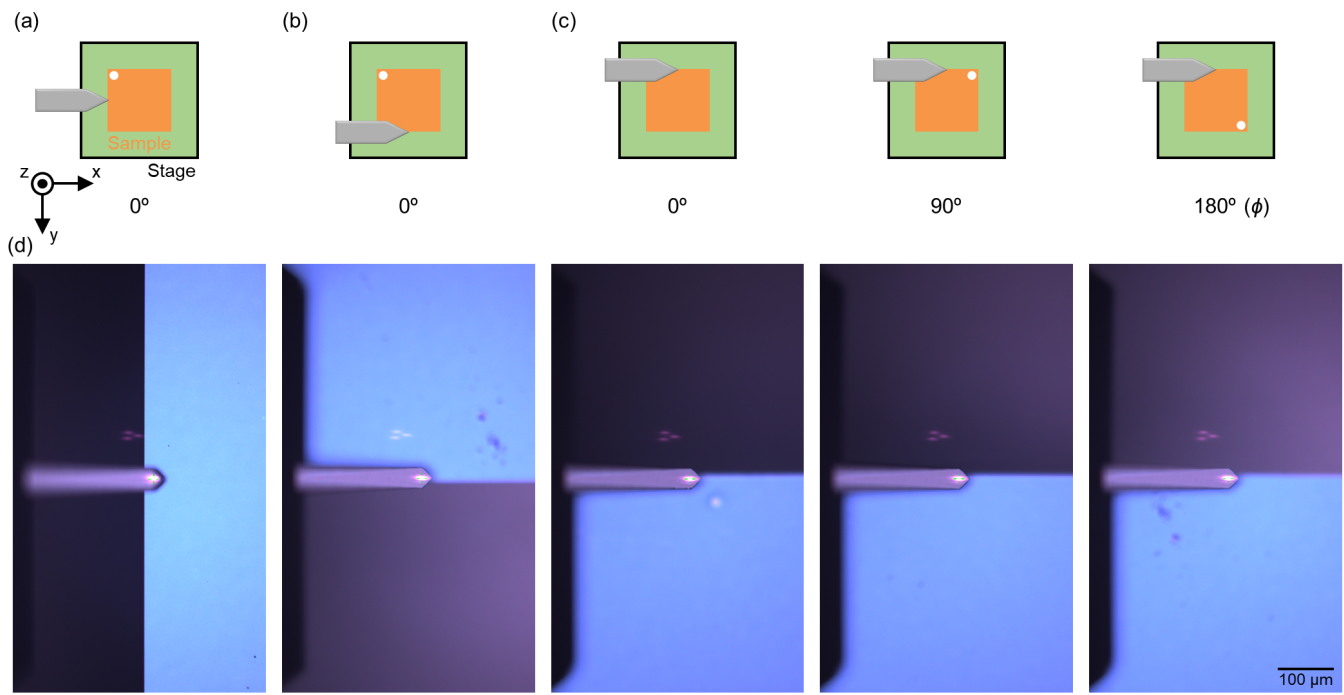

FIG S8. Schematic illustrations of a cantilever and a sample in PFM scanning where the long axis of the cantilever is perpendicular (a) or parallel (b) and (c) to the sample edge as a function of sample rotation angle ($\phi$). (d) Corresponding optical microscope images of the cantilever and the sample of Fig 6.

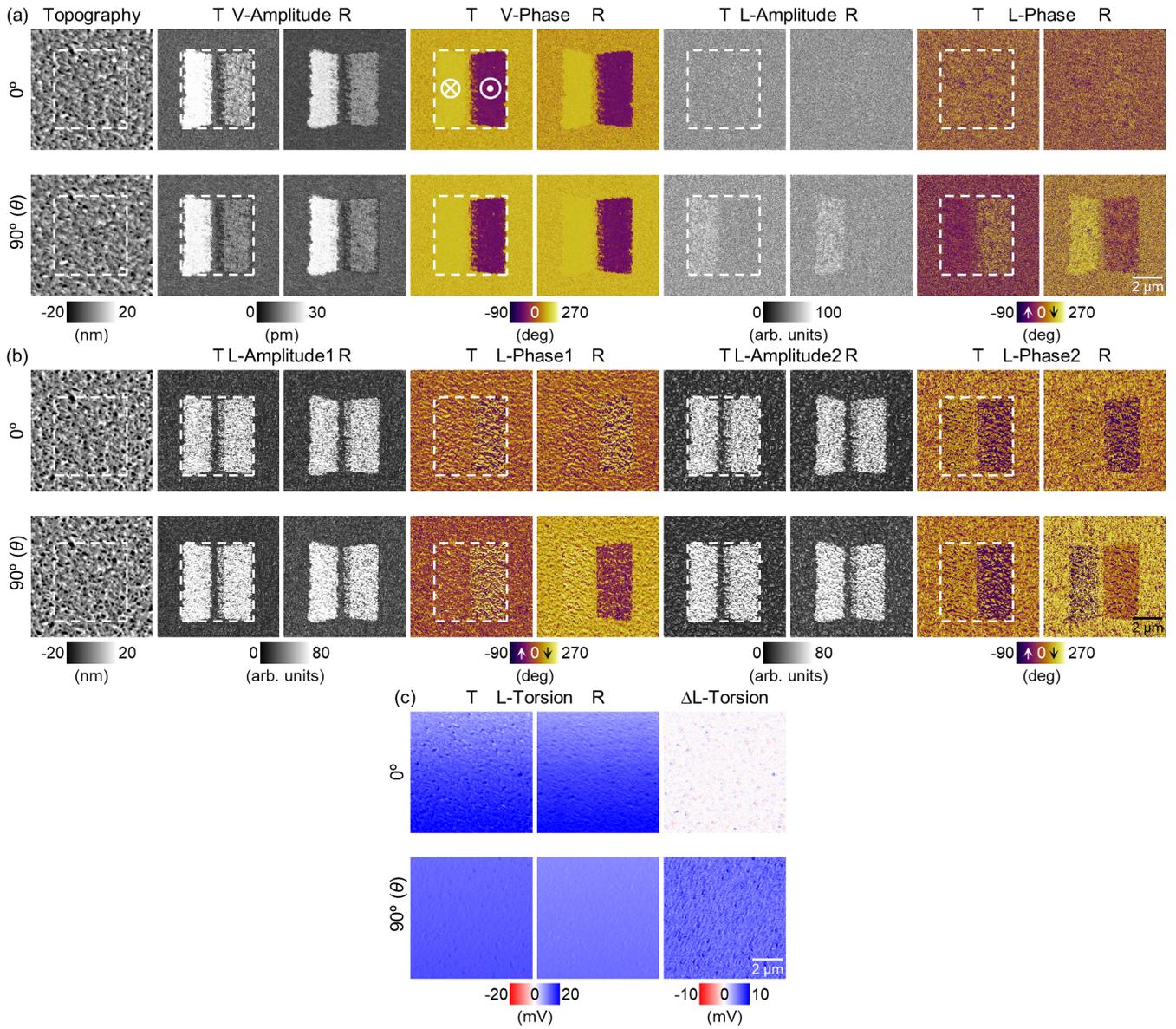

FIG S9. Topography, VPFM and LPFM images obtained by off-resonance PFM mode (a) and DART-LPFM mode (b) with a scan angle ($\theta$) of 0° and 90° on up and down poled domains (denoted by dash frames) in a P(VDF-TrFE) thin film created by the same procedure in Fig. 1. (c) Lateral torsion images and the corresponding discrepancy ($\Delta$) maps calculated by subtracting the retrace scan images from the trace scan images (T-R) obtained with a scan angle ($\theta$) of 0° and 90°. The lateral torsion was measured simultaneously with the off-resonance PFM.